# Kinetics of Stacking Order Evolution During Heterogeneous Ice Formation

Xudan Huang[1,2], Zifeng Yuan[3], Chon-Hei Lo[3], Huacong Sun[1,2], Lei Liao[1,2], Hongbo Han[1,2], Wenxi Li[1,4], Wenlong Wang[1,2], Zhi Xu[5], Lei Liu[4,6], Xuedong Bai[1,2,*], Limei Xu[3,†], Enge Wang[1,3,5,6,‡], Lifen Wang[1,2,§]

[1]Beijing National Laboratory for Condensed Matter Physics and Institute of Physics, Chinese Academy of Sciences, Beijing, 100190, China

[2]School of Physical Sciences, University of Chinese Academy of Sciences, Chinese Academy of Sciences, Beijing, China

[3]School of Physics, Peking University, Beijing 100871, China.

[4]Department of Materials Science and Engineering, College of Engineering, Peking University, Beijing 100871, P. R. China

[5]Tsientang Institute for Advanced Study, Zhejiang 310024, China.

[6]Interdisciplinary Institute of Light-Element Quantum Materials and Research Center for Light-Element Advanced Materials, Peking University, Beijing 100871, China.

The selection of stacking order in a broad range of close-packed polymorphic materials remains a challenging enigma. Using *in situ* cryogenic transmission electron microscopy, we uncover the atomistic mechanisms governing the vapour deposition growth of ice. We find that the heterogeneous ice nucleation and growth undergoes recrystallization accompanied by bifurcation, reflecting a coherent epitaxial transition from a cubic-ice embryonic core to hexagonal-ice prismatic dendrites, with intermediate stacking-disordered layers serving as a dynamic fluctuating bridge. Supported by molecular dynamics simulations, these phenomena are attributed to a surface-constrained, symmetry-breaking crystallization preference aligned with the principle of minimizing free energy. Our results highlight the critical role of the combined effects of surface and symmetry in shaping ice crystallization, providing fresh insights into crystal growth mechanisms and guiding principles for the design of advanced materials.

Despite its ubiquity, the crystallization of water ice remains enigmatic [1]. Ice I, the first identified ice polymorph, exists in two forms differing only in stacking sequences of water layers: hexagonal-close packed ice Ih [2] and cubic close-packed ice Ic [3-7]. Although ice Ih is more prevalent, both polymorphs can coexist up to ~ 240 K [8-15], and the rare 22° halo (Scheiner's Halo) has been proposed as evidence for atmospheric ice Ic [6,16]. Upon heating, ice Ic irreversibly transforms into ice Ih, with an estimated free-energy difference of approximately 30 to 50 J mol$^{-1}$ [11,17,18]. These subtle energetic and structural distinctions raise fundamental questions about the microscopic mechanism of ice growth.

Macroscopic observations of natural snow polycrystals suggest that ice Ic embryos may form during early crystallization, potentially explaining the characteristic ~70.5° inter-dendritic angle [19-23]. Yet, direct experimental and theoretical evidence remains scarce, hindered by limited temporal and spatial resolution [24].

An alternative view invokes stacking disordered ice (Isd) [25], comprising random sequences of cubic and hexagonal layers. Molecular dynamics (MD) simulations using the monoatomic water (mW) model, which accurately reproduces the melting point, suggest that structural entropy can stabilize ice Isd [26,27], possibly making it the embryonic nucleating phase. *In situ* cryogenic transmission electron microscopy (cryo-TEM) reveals a preference for ice Ic nucleation during vapour deposition [15]. Whether ice Isd is intrinsically more stable than ice Ih, or instead emerges from non-equilibrium dynamical effects associated with competing stacking orders, however, remains an open question.

In this study, we employed *in situ* cryo-TEM to investigate stacking-order selection during ice growth under far-from-equilibrium conditions at low temperature (~102K) and low pressure (~10$^{-6}$Pa). Using a metal sheet as the cold wall and molecular-resolution TEM imaging [15,28] to track vapour-deposited ice growth from the substrate toward the vacuum, we captured the coupled evolution of stacking order and morphology, from an embryonic ice Ic hemisphere, through stacking disordered intermediates, to ice Ih prismatic branches along the close-packed direction. Our findings reveal the surface-constrained metastability of cubic stacking and underscores the critical role of dynamically disordered stacking as a transient intermediate that mediates the transition from cubic to hexagonal order.

Figure 1 presents *in situ* cryo-TEM observations viewed along the direction perpendicular to ice growth. The experimental setup is schematically illustrated in Fig. 1 (a) (see details in Supplemental Material [29]). When the metal substrate (a thin sheet of copper or gold) serving as the cooling wall on the TEM holder was gradually cooled from room temperature to 102 K (Fig. S1), residual vapour in the TEM column vacuum (~$10^{-6}$ Pa) began to condense and deposit on its surface. *In situ* electron energy loss spectroscopy (EELS) confirmed the formation of water ice (Fig. S2). Individual nuclei and crystallites were subsequently tracked from the earliest stages of nucleation and growth using high-resolution *in situ* TEM imaging.

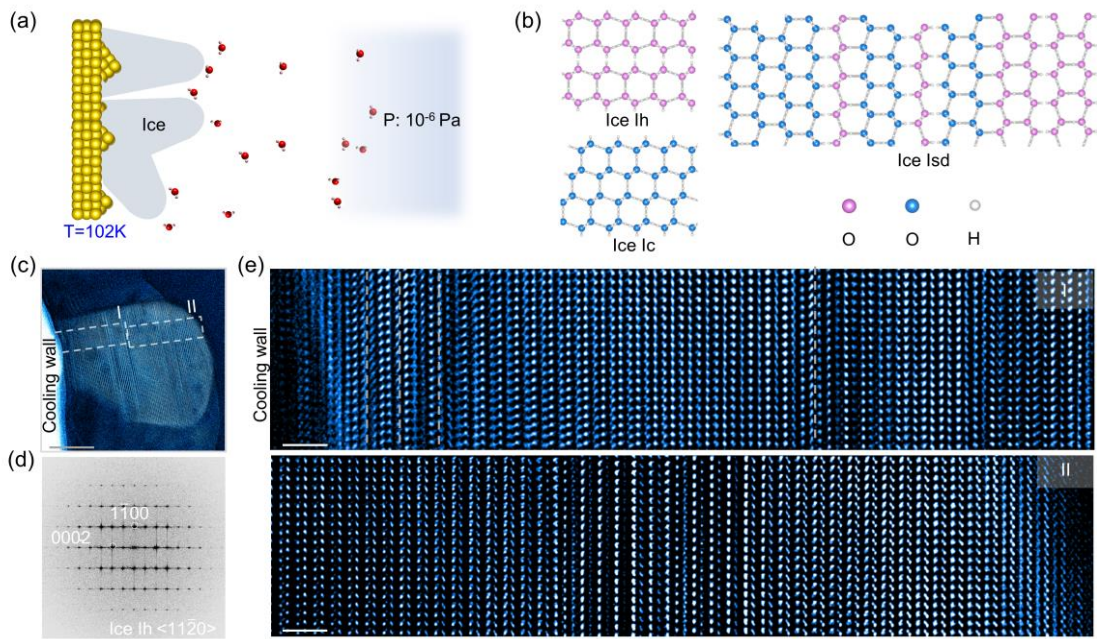

FIG. 1. Molecular-resolution TEM imaging of the stacking order in ice I. (a) Schematic of heterogenous ice nucleation and growth on cryogenic walls through vapour deposition in a TEM vacuum environment. (b) Structure illustration of polymorphic ice I. (c), (d) TEM image and the corresponding indexed diffractogram respectively of an ice Ih crystallite oriented along the <11$\bar{2}$0> crystallography direction. (e) High-resolution TEM micrograph of the crystallite in (c), showing the hexagonal close-packed stacking at molecular resolution, stacking faults are marked by dashed lines. Scale bars: 20 nm in (c); 2 nm in (e).

The schematic illustrations in Fig. 1(b) depict the stacking configurations of water-layer arrangements in the ice I polytypes. Ice Ih exhibits mirror symmetry along the stacking direction, whereas ice Ic displays centrosymmetry. A random sequence of mirror-symmetric hexagonal and centrosymmetric cubic layers gives rise to ice Isd.

Figure 1(c) shows a low-magnification TEM micrograph of an ice deposit, where lattice fringes extended continuously from the core to the tip of the crystallite. The corresponding indexed diffractogram identifies an ice Ih structure growing along the <0001> direction, viewed along the <11$\bar{2}$0> crystallography axis (Fig. 1d). A zoomed-in micrograph with molecular resolution (Fig. S3), further confirms hexagonal stacking along the growth direction, with four stacking faults detected among the 196 resolved layers (Fig. 1e). This columnar crystallite, exposing {0001} and {10$\bar{1}$0} facets of the hexagonal prism, represents the nucleation and growth of ice Ih, which, albeit thermodynamically stable, accounts for only a minority of the observed deposits.

Most deposits undergo a coupled morphological and structural transition from a hemispherical Ic core to needle-like or branched Ih dendrites (Fig. 2). Overview TEM micrographs (Figs. 2a–d) show distinct cross striations perpendicular to the growth direction, marking the progressive evolution from hemispherical to dendritic forms. The branching angle of ~ 70.5° coincides with the crystallographic angle between equivalent close-packed {111} planes of Ic core, indicating preferential growth along these planes. This directional preference drives dendrite bifurcation and supports the ice Ic embryo hypothesis inferred from macroscopic snow polycrystal observations [19,21,23].

High-resolution TEM micrographs in Fig. 2(b) reveal that the hemispherical cores of the deposits consist of cubic stacking layers. The pronounced striations above the core exhibit stacking-disordered sequences, characterized by an increasing proportion of hexagonal layers and a decreasing proportion of cubic layers, all aligned with the equivalent close-packed {111} planes of the Ic core, indicating coherent epitaxial growth. At the growth front, the branch tip is composed entirely of hexagonal stacking layers.

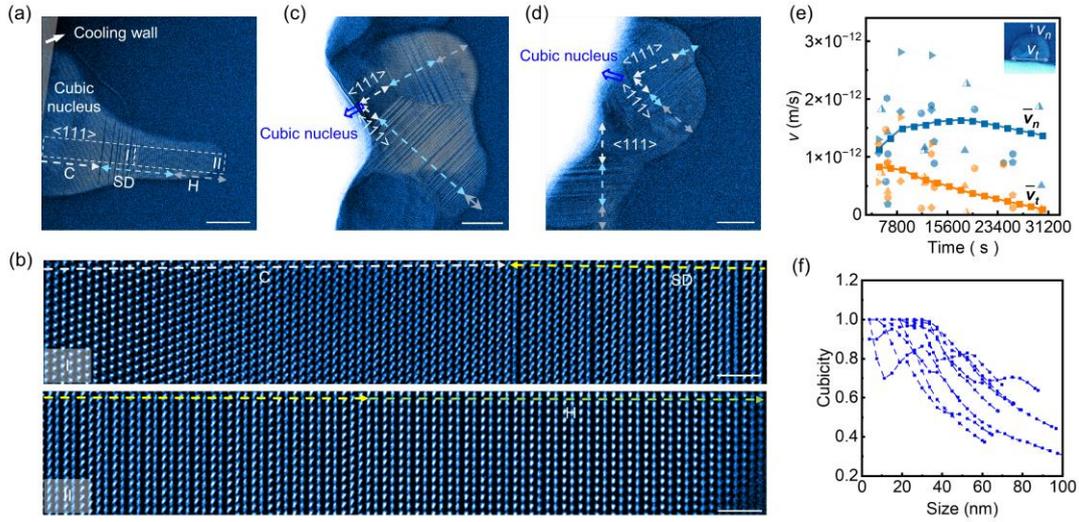

FIG. 2. The growth evolution of stacking orders in ice I. (a) TEM micrograph of a grown ice, illustrating morphology transition along the growth direction. (b) High-resolution TEM micrograph of the crystallite in (a), depicting the evolution of stacking orders from cubic (C) to stacking disorder (SD), and finally to hexagonal (H), progressing from the base to the tip along the <111> layer stacking direction of the cubic nucleus. (c), (d) TEM micrographs of grown crystallites, demonstrating the bifurcation of a hemispherical cubic ice nucleus into hexagonal dendritic branches, with intervening stacking-disordered structures along equivalent <111> water layer stacking directions of cubic ice. (e) Statistics of growth velocities of in plane ($v_t$), and out of plane ($v_n$), showing the transition from isotropic to anisotropic growth from the core to the branches. (f) Size-dependent cubicity, defined as the percentage of cubic stacking layers within the overall stacking layers of ice branches, indicating a gradual transition from cubic to hexagonal stacking growth. Scale bars: 20 nm in (a), (c), and (d); 2 nm in (b).

The statistics regarding the growth velocities of the observed ice crystallites (Fig. 2e), further elucidate the evolution of growth modes. Initially, isotropic growth dominates; progressively, anisotropic growth develops and increasingly favors out-of-plane directions at later stages. This transition suggests a symmetry-breaking mechanism underlying the growth transformation from an ice-Ic core (four equivalent {111} planes and corresponding <111> directions) to an ice-Ih prismatic branch with a single close-packed stacking axis, aligned with the stacking direction, consistent with the principle of free-energy minimization during phase evolution. Concurrently, the structural cubicity—defined as the fraction of cubic layers within the stacking sequences—decreases from the base to the tip, as shown in Fig. 2(f).

The ice Ic core, approximately 20 nm in diameter, were repeatedly observed exposing equivalent close-packed planes that bifurcate into Ih dendritic branches (see also Fig. S4–S7). This heterogeneous nucleation preference is consistent with previous reports on the metastable nature of ice Ic relative to ice Ih and adheres to Ostwald's step rule of phase evolution [30], which describes transitions from metastable to stable states through intermediate phases. The experimentally observed stacking-disordered layers bridging the Ic core and the Ih branches reflect a dynamic interface, where local stacking fluctuations facilitate the cubic-to-hexagonal transformation, revealing a non-equilibrium pathway through which ice reorganizes its stacking sequence during growth.

Furthermore, the role of coherent epitaxial stacking in facilitating phase-transformation growth is corroborated by statistical observations showing that coupled structural and morphological evolution occurs only when the growth direction aligns with the layer-stacking axis. Pure-phase ice Ic, as shown in Figure 3, forms when growth proceeds off the <111> axis of the ice Ic core. In Fig. 3(a–c), three ice crystallites growing along directions away from <111>, thus lacking a coherent axis with ice Ih, develop into single crystalline Ic structure free of significant defects. Crystallographic statistics for all observed pure-phase ice Ic (Fig. 3d) reveal varying growth directions other than <111>, with sizes up to 110 nm, comparable to single crystalline ice Ih in its equilibrium state (Fig. S8). This size is substantially larger than the threshold for the cubic-to-hexagonal transition along <111>, implying successful kinetic trapping of the metastable phase, enabled by surface protection that preserves epitaxial integrity and prevents symmetry-breaking transitions. This methodology may be readily extended to the controlled design of single crystals in other polymorphic systems.

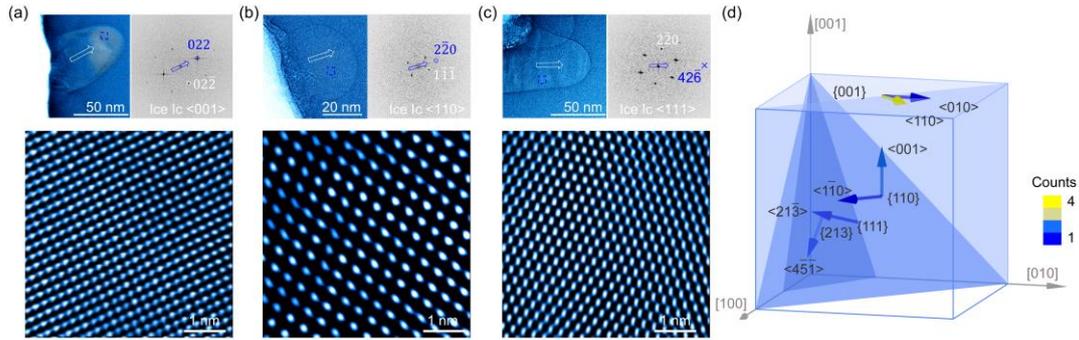

FIG. 3. Pure-phase ice Ic. (a)–(c) Cubic ice crystallites viewed along the <001>, <110>, and <111> directions, respectively, demonstrating that ice growth away from the <111> stacking direction of the water layers yields a single-crystalline cubic phase. (d) Statistical analysis of the imaging projection planes and growth orientations of the observed pure-phase cubic ice crystallites.

Previous theoretical studies report negligible differences in thermodynamically relevant qualities, such as free energy and nucleation barriers, among ice polytypes [25,31-34]. Yet, mW-model simulations predicts that ice Isd is stabilized by entropic effects, dominating homogeneous nucleation along the classical nucleation pathway [26]. This contradicts heterogeneous simulations suggesting multiple nucleation routes for ice Ic [35], and the Ostwald's step rule-based assumption of cubic ice embryo formation [19,20]. Moreover, the dynamic behavior of heterogeneous ice growth and the accompanying stacking-order evolution observed in our experiments cannot be fully reconciled with either one-step or two-step nucleation and growth theories. To gain a more comprehensive understanding of these structural and kinetic processes, we have performed MD simulations of vapour-deposition growth of ice [35-42] (see details in Supplemental Material[29]), as shown in Fig. 4.

Given the negligible bulk energy difference between Ic and Ih [43], we compared the coarse-grained surface energies ($\gamma$) of ice nuclei with hemispherical and branch-like morphologies (Fig. 4a, Fig. S10). Branch-like Ih is most stable, followed by hemispherical Ic, with hemispherical Ih and branch-like Ic less favorable for finite-sized clusters. Calculations using a machine learning force field for water model [44] (Fig. S11) corroborate this trend, suggesting that the experimentally observed formation of hemisphere Ic cores and Ih prism branches may be driven by morphology-dependent surface energy preferences.

By designing ice nuclei with controlled morphologies to evaluate free energy as a function of branch growth on an ice Ic core (Fig. 4b), we find that the initial stages of branch formation show negligible energetic differences among Ic, Isd, and Ih. This indicates that stacking-disordered structures are not energetically disfavored during early rcrystallization. As the branch grows, however, the Ih structure becomes increasingly preferred.

MD simulations of water vapor deposition on an ice Ic core reveal the growth of stacking-disordered layers, consistent with our experimental observations. Representative snapshots of the deposition trajectories (inset, Fig. 4c) show diverse coordination with the top surface and pronounced structural fluctuations reflecting competition between Ic and Ih stacking [43]. This leads to random sequences of cubic and hexagonal layers, generating stacking faults that mediate the transformation from a hemispherical Ic core to an Ih prism branch (Fig. S12). The statistical cubicity of the crystallites decreases nonlinearly with size along the stacking direction (Fig. 4c), confirming that intermediate stacking disorder bridges the surface–constrained metastable Ic core and the thermodynamically stable Ih prism.

The competition between ice Ic and Ih has long been explored through hydrogen-bond residual entropy [3], and mW-model simulations [26], all suggesting that stacking selection is inherently indeterminate, with ice Ic remaining metastable within the hexagonal phase space [5,11,13,45].

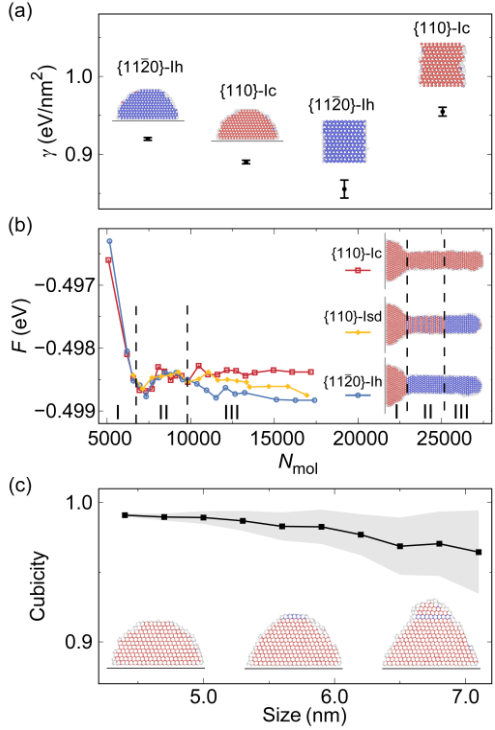

FIG. 4. Energetics and growth dynamics of ice polytypes. (a) Surface energies ($\gamma$) of finite ice clusters, showing that branch-like ice Ih and hemisphere ice Ic are more stable than hemisphere Ih or branch-like ice Ic. (b) Free energy as a function of molecule number $N_{mol}$, indicating neglectable energetic differences among the polytypes (Ic, Isd, Ih) during early growth on an ice Ic core, followed by a progressive stabilization of the Ih branch at later stages. (c) Statistical cubicity of MD-simulated vapor-deposited-ice on an ice Ic core as a function of crystallite size along the stacking direction. The black line and grey shading denote the mean and standard deviation from ten independent MD trajectories. Insets show representative MD snapshots of the deposition process.

Our results elucidate the surface–constrained metastability of cubic stacking during heterogeneous ice nucleation and growth, corroborating the polymorphic nature of the ice I phase space. Moreover, we demonstrate that pure-phase, single-crystalline ice Ic can be synthesized via off-axis deposition growth, highlighting a general strategy for stabilizing surface-confined, high-symmetry metastable structures in other materials.

Furthermore, our study depicts an atomistic picture of an intrinsically non-equilibrium vapor–solid phase transition in water, revealing how the interplay of symmetry and surface dictates heterogeneous ice crystallization kinetics and thermodynamics. The direct visualization of heterogeneous ice nucleation and subsequent fluctuation-driven bifurcation-recrystallization under cryogenic vacuum conditions offers both a mechanistic foundation for snowflake morphogenesis and fresh insight into the microscopic origins of the macroscopic quantum phase transitions in water.

This work was supported by the National Key R&D Program of China (2024YFA1208201, 2021YFA1400500, 2021YFA1400204), National Natural Science Foundation of China (52322311, 52461160301, 52427802, 12334001, 11935002), and the Postdoctoral Fellowship Program of CPSF (GZB20240813). Numerical simulations were carried out on the TianHE-1A supercomputer. L.W. is grateful for the support from the Youth Innovation Promotion Association of CAS (2020009).

X.H. and Z.Y. contributed equally to this work.

*xdbai@iphy.ac.cn;

†limei.xu@pku.edu.cn;

‡egwang@pku.edu.cn;

§wanglf@iphy.ac.cn;